\documentclass[sigconf]{acmart}

\AtBeginDocument{%
  }

\copyrightyear{2024}
\acmYear{2024}
\setcopyright{acmlicensed}
\acmConference[SIGIR '24]{Proceedings of the 47th International ACM SIGIR Conference on Research and Development in Information Retrieval}{July 14--18, 2024}{Washington, DC, USA}
\acmBooktitle{Proceedings of the 47th International ACM SIGIR Conference on Research and Development in Information Retrieval (SIGIR '24), July 14--18, 2024, Washington, DC, USA}
\acmDOI{10.1145/3626772.3657690}
\acmISBN{979-8-4007-0431-4/24/07}

\newcommand{\ie}{\emph{i.e., }}
\newcommand{\eg}{\emph{e.g., }}

\usepackage{subcaption}
\usepackage{multirow,multicol}
\usepackage{balance}
\usepackage{enumitem}

\begin{document}

\title{LLaRA: Large Language-Recommendation Assistant}

\author{Jiayi Liao}
\email{ljy0ustc@mail.ustc.edu.cn}
\orcid{0009-0006-7998-8462}
\affiliation{%
  \institution{University of Science and Technology of China}
  \city{Hefei}
  \country{China}
}

\author{Sihang Li}
\email{sihang0520@gmail.com}
\orcid{0009-0009-8986-7965}
\affiliation{%
  \institution{University of Science and Technology of China}
  \city{Hefei}
  \country{China}
}

\author{Zhengyi Yang}
\email{yangzhy@mail.ustc.edu.cn}
\orcid{0009-0009-8094-0978}
\affiliation{%
  \institution{University of Science and Technology of China}
  \city{Hefei}
  \country{China}
}

\author{Jiancan Wu}
\email{wujcan@gmail.com}
\orcid{0000-0002-6941-5218}
\affiliation{%
  \institution{University of Science and Technology of China}
  \city{Hefei}
  \country{China}
}

\author{Yancheng Yuan}
\email{yanchengyuanmath@gmail.com}
\orcid{0000-0002-8243-4683}
\affiliation{%
  \institution{The Hong Kong Polytechnic University}
  \city{Hong Kong}
  \country{China}
}

\author{Xiang Wang}
\authornote{Corresponding authors, and they are also affiliated with Institute of Dataspace, Hefei Comprehensive National Science Center.}
\email{xiangwang1223@gmail.com}
\orcid{0000-0002-6148-6329}
\affiliation{%
  \institution{University of Science and Technology of China}
  \city{Hefei}
  \country{China}
}

\author{Xiangnan He}
\authornotemark[1]
\email{xiangnanhe@gmail.com}
\orcid{0000-0001-8472-7992}
\affiliation{%
  \institution{University of Science and Technology of China}
  \city{Hefei}
  \country{China}
}

\renewcommand{\shortauthors}{Jiayi Liao et al.}

\begin{abstract}
Sequential recommendation aims to predict users' next interaction with items based on their past engagement sequence.
Recently, the advent of Large Language Models (LLMs) has sparked interest in leveraging them for sequential recommendation, viewing it as language modeling.
Previous studies represent items within LLMs' input prompts as either ID indices or textual metadata. 
However, these approaches often fail to either encapsulate comprehensive world knowledge or exhibit sufficient behavioral understanding.
To combine the complementary strengths of conventional recommenders in capturing behavioral patterns of users and LLMs in encoding world knowledge about items, we introduce \textbf{L}arge \textbf{La}nguage-\textbf{R}ecommendation \textbf{A}ssistant (\textbf{LLaRA}). 
Specifically, it uses a novel hybrid prompting method that integrates ID-based item embeddings learned by traditional recommendation models with textual item features.
Treating the ``sequential behaviors of users'' as a distinct modality beyond texts, we employ a projector to align the traditional recommender's ID embeddings with the LLM's input space. 
Moreover, rather than directly exposing the hybrid prompt to LLMs, a curriculum learning strategy is adopted to gradually ramp up training complexity. 
Initially, we warm up the LLM using text-only prompts, which better suit its inherent language modeling ability. 
Subsequently, we progressively transition to the hybrid prompts, training the model to seamlessly incorporate the behavioral knowledge from
the traditional sequential recommender into the LLM.
Empirical results validate the effectiveness of our proposed framework.
Codes are available at \url{https://github.com/ljy0ustc/LLaRA}.
\end{abstract}


  


\maketitle

\section{Introduction}
Sequential recommendation \cite{seq_rec_survey1, seq_rec_survey2} is to predict users' next items of interest based on their historical interactions with items.
Conventional sequential recommenders \cite{GRU4Rec,Caser,SASRec} typically involve two steps:
(1) assigning each item with a distinct ID, which is converted into a trainable embedding; 
(2) learning these embeddings with the objective of next item prediction, so as to capture user preference.
After training on historical interaction data, item representations can encapsulate the sequential behavioral patterns of users.

\begin{figure*}
    \centering
    \begin{subfigure}[b]{0.25\textwidth}
        \includegraphics[width=\textwidth]{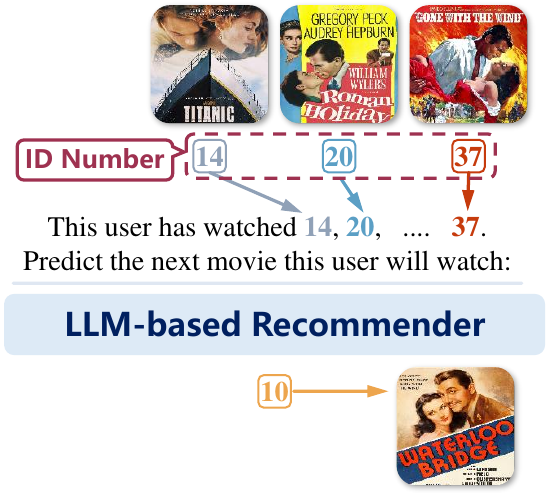}
        \caption{ID Number}
        \label{fig1:sub_a}
    \end{subfigure}
    \begin{subfigure}[b]{0.32\textwidth}
        \includegraphics[width=\textwidth]{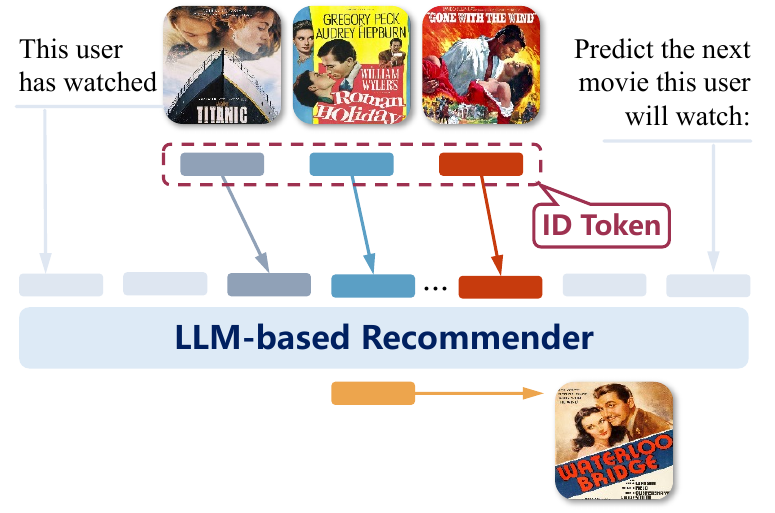}
        \caption{Randomly-initialized ID Token}
        \label{fig1:sub_b}
    \end{subfigure}
    \begin{subfigure}[b]{0.3\textwidth}
        \includegraphics[width=\textwidth]{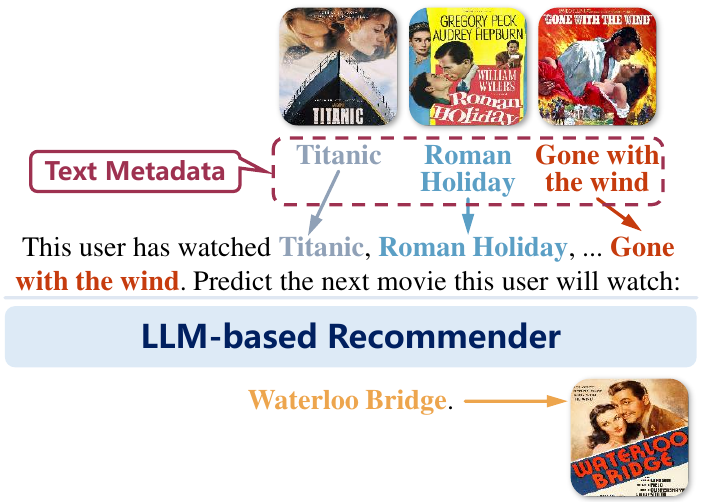}
        \caption{Text Metadata}
        \label{fig1:sub_c}
    \end{subfigure}
    \begin{subfigure}[b]{0.72\textwidth}
        \includegraphics[width=\textwidth]{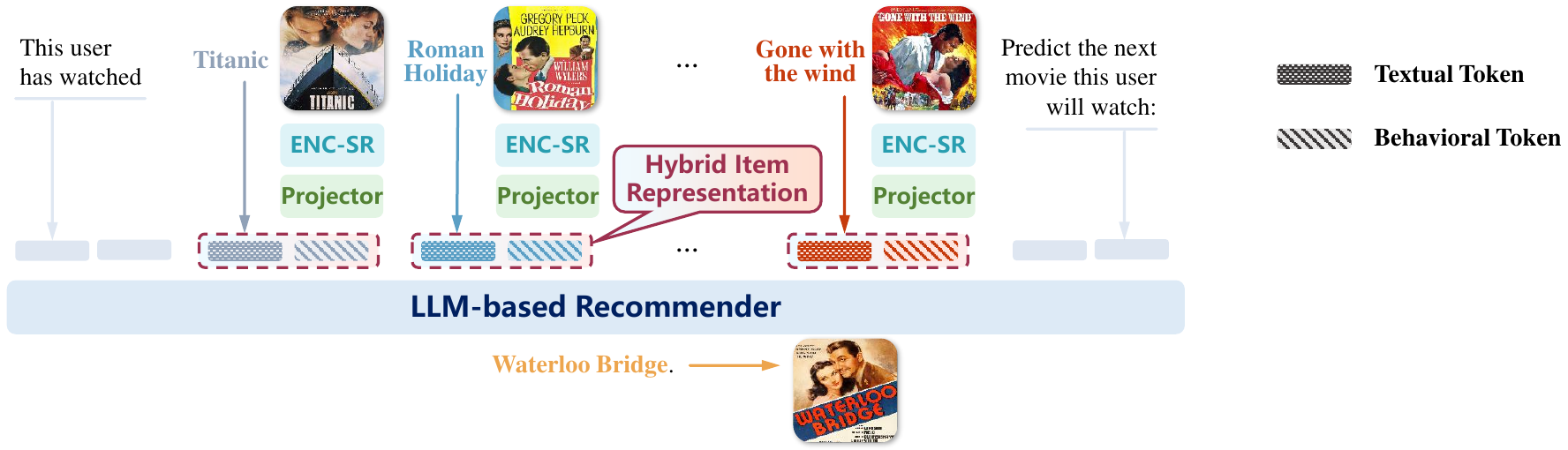}
        \caption{Hybrid Item Representation}
        \label{fig1:sub_d}
    \end{subfigure}
    \vspace{-5pt}
    \caption{Comparison among three prior item representation methods and ours. (a) ID Number: represents an item with a numerical index. (b) Randomly-initialized ID Token: represents an item with an OOV-independent token. (c) Text Metadata: represents an item with its textual features, such as item title. (d) Hybrid Item Representation: integrates both textual tokens and behavioral tokens derived from the ID-based item embedding learned by traditional recommender models.}
    \label{fig1}
    \vspace{-5pt}
\end{figure*}

Recently, inspired by the great success of Large Language Models (LLMs) \cite{gpt3, llama, vicuna}, exploring the potential of LLMs in sequential recommendation is attracting attention \cite{RecFormer, TALLRec, P5, M6-Rec, uncovering_ChatGPT4Rec, is_chatgpt_a_good_recommender, LLM-as-zero-shot-rankers,agent4rec,ToolRec}, especially driven by extensive world knowledge and innate reasoning capabilities of LLMs.
At the core is to reshape sequential recommendation as the language modeling task --- that is, convert the behavioral sequence into the textual input prompt, \eg ``This user has watched [$\texttt{item}_1$], [$\texttt{item}_2$], \ldots, [$\texttt{item}_n$]. Predict the next movie this user will watch.''.
When considering the way to represent the item within the prompt (\eg [$\texttt{item}_k$]), prior studies generally follow two approaches:

\begin{itemize}[leftmargin=*]

\item ID-based Representation: 
Within the prompt, each item is represented as an ID number \cite{P5} (\eg ``14'' for the movie ``Titanic'') or a randomly-initialized ID token \cite{P5-index}, as Figures \ref{fig1:sub_a} and \ref{fig1:sub_b} illustrate, respectively.
Despite its simplicity, this approach leaves the textual characteristics of items (\eg titles and descriptions) untouched, consequently underutilizing the world knowledge inherent in LLMs.
Moreover, the employment of ID numbers or ID tokens might pose integration challenges with LLMs, as it does not correspond well with the natural language processing capabilities of LLMs.
    
\item Text-based Representation: 
This approach encodes each item in the prompt through its textual metadata, such as titles \cite{TALLRec, M6-Rec} and descriptions \cite{RecFormer, VQ-Rec}.
Taking Figure \ref{fig1:sub_c} as an example, the movie can be directly represented by its title ``Titanic''.
While effectively harnessing LLMs' linguistic capabilities and world knowledge about items, it falls short of exhibiting the sequential behavior patterns of users.
Overlooking such patterns could confine LLM in a suboptimal position when predicting the next item.
\end{itemize}
Consequently, we argue that merely prompting LLMs with either ID-based or text-based representations of item sequences fails to fully tap into LLMs' potential for sequential recommendation.
Instead, the LLMs should gain a deeper understanding of the behavioral patterns inherent in the sequential interactions.

In pursuit of this goal, we explore the alignment between LLMs and the sequential recommenders, going beyond relying on mere ID-based or text-based prompting.
Drawing inspiration from Multi-modal Large Language Models (MLLMs) \cite{Flamingo, minigpt4, PALM-E, AudioGPT} that adeptly understand and reason across diverse modalities (\eg images, audio, and 3D point clouds), we propose viewing the ``sequential behaviors of users'' as a new modality for LLMs in recommendation and aligning it with the language space.
Such an alignment could empower LLMs to understand and internalize the behavioral patterns that recommenders have effectively identified and utilized.

To this end, we propose a novel framework as illustrated in Figure \ref{framework}, named \textbf{L}arge \textbf{La}nguage-\textbf{R}ecommendation \textbf{A}ssistant (\textbf{LLaRA}), which integrates conventional sequential recommenders into LLMs with two tailor-made enhancements:

\textbf{(1) Hybrid Prompt Design:}
We exploit two distinct approaches, text-only and hybrid prompting, to convert an interaction sequence into an input prompt for LLMs.
Specifically, the text-only method represents each item using its textual metadata, which are then transformed into textual tokens.
Beyond text-only prompting, we further devise hybrid prompting, which integrates behavioral patterns sourced from recommenders.
That is, for an item's ID representation from a traditional recommender (\eg SASRec \cite{SASRec}), we feed it into a projector (\eg a trainable MLP)
to yield a behavioral token that is compatible with the LLMs' textual token space.
We then combine the textual and behavioral tokens, creating a multifaceted representation of each item within the prompt.
Considering the movie example in Figure \ref{fig1:sub_d}, [$\texttt{item}_1$] is depicted as the concatenation of textual token \includegraphics[width=0.04\textwidth]{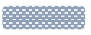} of word ``Titanic'' and behavioral token \includegraphics[width=0.04\textwidth]{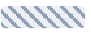}.
Such an integration offers a more holistic depiction of user behaviors, surpassing prompts solely based on the ID or text.

\textbf{(2) Curriculum Prompt Tuning:}
Building upon the dual prompting approaches, we draw inspiration from curriculum learning \cite{curriculm_learning, cl_survey} and propose a curriculum prompt tuning strategy --- gradually shifting the learning focus from text-only prompting to hybrid prompting.
Specifically, our strategy begins with text-only prompting, serving as an initial warm-up phase for the LLM.
This phase is designed to align with the natural language modeling capabilities of the LLM, as it involves characterizing items through their textual metadata.
The tuning in this phase ensures that the LLM becomes acquainted with the basic idea of the recommendation mechanism.
Following this, we transition to hybrid prompting, which trains the projector to inject behavioral knowledge from recommenders into the LLM effectively.

Overall, we not only familiarize the LLM with the recommendation mechanism utilizing text-only prompts, but also internalize the behavioral knowledge encoded by recommenders with hybrid prompts.
The progressive tuning strategy ensures an evolving learning experience for the LLM, enhancing its capabilities of sequential recommendation with a deeper understanding of user behavior.

We conduct experiments on three datasets, MovieLens \cite{movielens}, Steam \cite{SASRec}, and LastFM \cite{lastfm},  to compare LLaRA with various leading sequential recommender models and several LLM-based methods.
The results show that LLaRA consistently outperforms these baselines in terms of the HitRatio@1 metric, demonstrating its superiority.
Furthermore, we perform ablation studies to justify the importance of the two key components: hybrid prompting and curriculum prompt tuning.

In summary, our contributions can be concluded as follows: We propose a novel framework, LLaRA, to enhance LLMs with sequential recommenders. In LLaRA, we introduce a hybrid prompting method that integrates both world knowledge and behavioral patterns into item representations; and we conduct curriculum prompt tuning to achieve modality alignment. Comprehensive experimental results underscore the effectiveness of the LLaRA framework.

\section{Related Work}
In this section, we provide a literature review pertaining to Large Language Models, Multi-modal Large Language Models, and LLMs for Sequential Recommendation. 
Our work draws inspiration from them for fusing LLMs and sequential recommendation systems.

\subsection{Large Language Models}
Language modeling has been extensively scrutinized for language understanding and generation over the past years, thereby catalyzing the recent emergence of Language Models (LMs) \cite{gpt3, T5, bert, llama, vicuna}. 
Pretrained LMs built on the Transformer architecture, such as BERT \cite{bert} and T5\cite{T5}, have demonstrated profound versatility owing to their large-scale training corpus. 
More recently, researchers have delved deeper into the scaling effect by augmenting the parameter and training corpus scale to an unprecedented magnitude --- encompassing billions of parameters and trillions of training tokens. 
These Large Language Models (LLMs), like GPT-4 \cite{GPT-4} and Llama \cite{llama}, manifest substantial performance enhancements and show emergent abilities, such as commonsense reasoning and instruction following.
Moreover, domain-specific LLMs, such as those in the domain of finance \cite{BloombergGPT}, medicine \cite{Med-PaLM}, and law \cite{chatlaw}, are constructed by integrating domain expertise with the commonsense knowledge inherent in general LLMs. 
These advancements inspire us to probe the potential of LLMs in the domain of recommendation.

\subsection{Multi-Modal Large Language Models}
Despite their versatility and promising performance, most LLMs are restricted to textual inputs. 
However, a vast reservoir of information and knowledge resides in other modalities, including vision, video, and audio. 
Consequently, researchers have proposed Multi-modal Large Language Models (MLLMs), to integrate the text with other modalities \cite{clip,declip}.
Recent MLLMs suggest that visual space can be harmoniously aligned with textual space \cite{blip2, minigpt4, Frozen, PALM-E}, thereby empowering them to perform language generation tasks conditioned on visual inputs.
Beyond vision, other modalities like video \cite{Video-LLaMA}, audio \cite{macawllm}, graph \cite{molca}, and 3D point clouds \cite{3D-LLM, 3d-molm} are incorporated into LLMs, enabling them to digest information and knowledge of other modalities. 
We draw inspiration from these prior studies to devise LLaRA, which fuses multi-modal information to enhance sequential recommendation.

\begin{figure*}
    \centering
    \begin{subfigure}[b]{0.6\textwidth}
        \includegraphics[width=\textwidth]{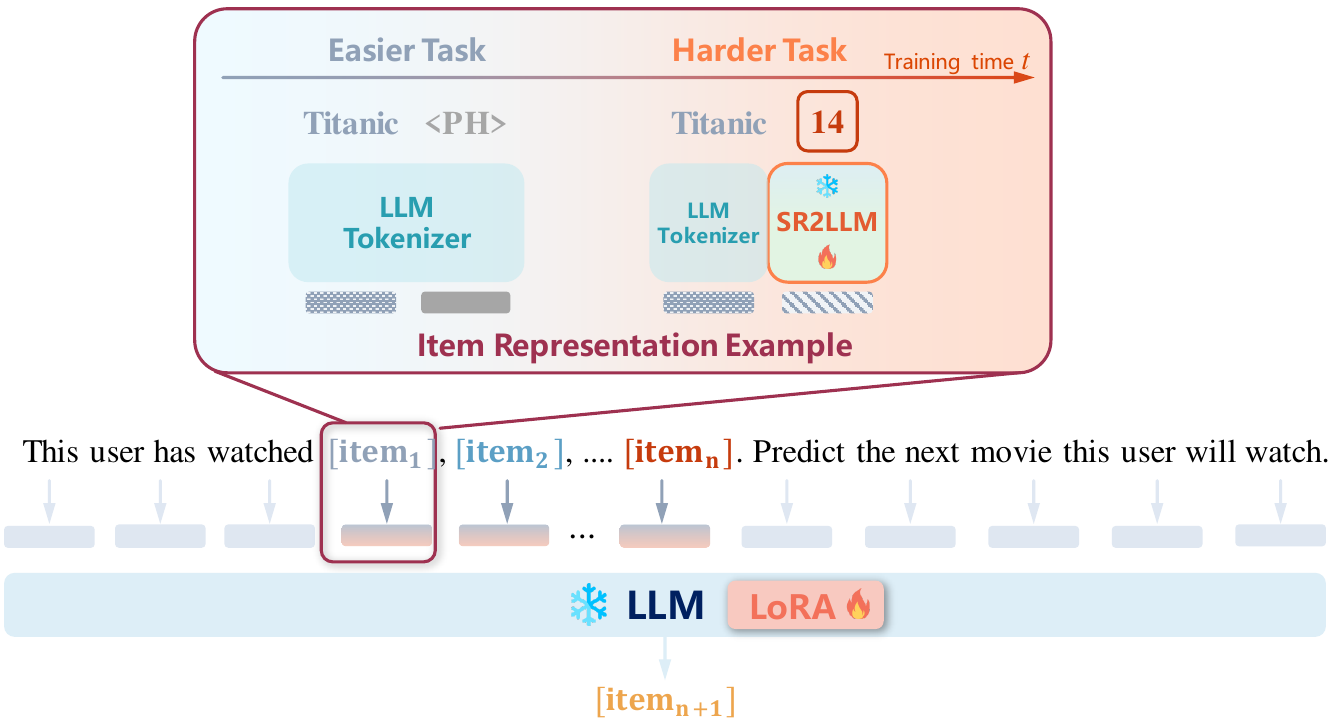}
        \caption{Curriculum Prompt Tuning for Hybrid Item Representation.}
        \label{framework}
    \end{subfigure}
    \begin{subfigure}[b]{0.3\textwidth}
        \includegraphics[width=\textwidth]{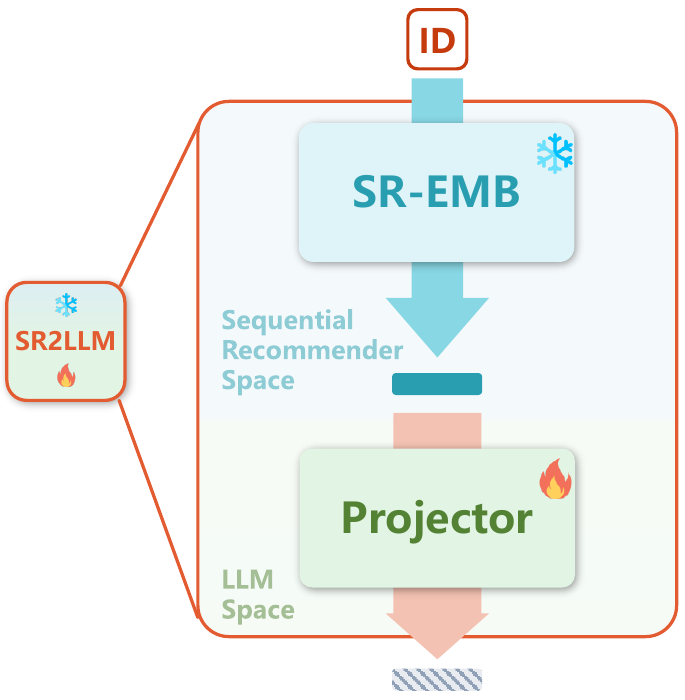}
        \caption{Architecture of SR2LLM.}
        \label{projector}
    \end{subfigure}
    \vspace{-6pt}
    \caption{The LLaRA framework. 
    (a) Sequential recommendation data is transformed into the instruction-tuning format. The item representation example illustrates the transition from pure textual tokens to the integration of the textual tokens with the behavioral token. (b) The sequential recommender is well-trained and frozen, while the trainable projector bridges the sequential recommender and LLM space.
    }
    \label{method}
    \vspace{-5pt}
\end{figure*}

\subsection{LLMs for Sequential Recommendation}
Sequential recommendation aims to predict the next item that matches user preference, based on his/her historical interaction sequence \cite{seq_rec_survey1,seq_rec_survey2}.
Prior studies have explored employing complex model architectures to better characterize user preference, including Recurrent Neural Networks (RNNs) \cite{GRU4Rec, RNN-based-SRS1, RNN-based-SRS2}, Convolutional Neural Networks (CNNs) \cite{Caser, CNN-based-SRS}, and Attention mechanisms \cite{SASRec, Bert4Rec}.
With the advent of LLMs, researchers pay increasing attention to exploring their potential for sequential recommendation.
Not only the extensive world knowledge stored in LLMs could serve as a rich source of background information for items \cite{GPT-4}, but also the reasoning capabilities of LLMs are able to augment the next item prediction \cite{llama2}.
When integrating LLMs into recommendation (LLM4Rec), there are two prevalent categories \cite{LLM4Rec-survey1, LLM4Rec-survey2}:
\begin{itemize}[leftmargin=*]
    \item \textbf{LLM as the Recommender.} It involves training from scratch \cite{RecFormer}, tuning \cite{TALLRec, P5, M6-Rec}, prompting \cite{uncovering_ChatGPT4Rec}, and in-context learning \cite{is_chatgpt_a_good_recommender, LLM-as-zero-shot-rankers} an LLM on recommendation data to serve as a recommender.
    Although studies within this category have substantiated that LLMs can be imbued with recommendation capabilities, they neglect established yet effective recommendation models.

    \item \textbf{LLM as the Enhancer.} It augments traditional recommenders with LLM tokens or embeddings \cite{MoRec_or_IDRec, VQ-Rec, UniSRec}.
    It typically utilizes LLMs as feature extractors or text generators, given their exceptional ability to integrate diverse sources and forms of information, such as item metadata.
    Nonetheless, the actual recommendation process is still done by conventional models, leaving the LLMs' reasoning skills untouched.
\end{itemize}

Different from the aforementioned studies, LLaRA investigates aligning traditional sequential recommendation models with LLMs. 
It not only capitalizes on the sequential behavioral patterns learned by the well-established recommender models, but also utilizes the reasoning ability and world knowledge embedded within LLMs.
In contrast to its concurrent work \cite{collm}, LLaRA introduces the curriculum tuning strategy to achieve this alignment, ensuring a more stable learning procedure, and concentrates on list-wise ranking instead of the point-wise binary (yes/no) classification task.
\section{Preliminary}

\textbf{Task Formulation.} 
Given a user who has chronologically engaged with item sequence $[i_1, i_2, \ldots, i_n]$, a sequential recommender entails predicting the next item  $i_{n+1}$ this user will interact with.

\vspace{4pt}\noindent\textbf{Curriculum Learning.}
Inspired by the pedagogical strategies in human education, curriculum learning \cite{curriculm_learning} emphasizes training the model from simpler to more complex learning tasks.
In general, it involves three critical stages \cite{cl_survey}:

(1) Complexity Assessment: This initial stage quantifies the complexity of each data point or task, which is then used to assign a learning priority. 

(2) Scheduler Formulation: Based on the complexity, a training scheduler is developed to arrange the sequence and frequency of tasks presented to the model, typically commencing with easier tasks and gradually advancing to harder ones.

(3) Training Execution: The curriculum learning process is implemented adhering to the predetermined progression. 

\vspace{4pt}\noindent\textbf{Instruction Tuning.}
Instruction tuning emerges as a pivotal approach that can substantially boost LLMs to follow human task-specific instructions \cite{instructGPT}.
Specifically, it first reorganizes data into $\mathcal{Z}=\{(x_i,y_i)\}_{i=1,..,N}$, where $x_i$ and $y_i$ denote the textual instructions and the corresponding responses respectively.
This pairing format not only encapsulates the task descriptions but also converts training data into a natural language format, thus creating a comprehensive instructional context.
Subsequently, we can tune the LLMs with $\mathcal{Z}$ following the autoregressive objective \cite{llama, gpt, gpt3}:
\begin{equation}
\max_\Phi\sum_{(x,y)\in\mathcal{Z}}\sum_{t=1}^{|y|}\log (P_\Phi(y_t|x,y_{<t})),
\end{equation}
where $\Phi$ is the parameters of the LLMs, with $y_t$ referring to the $t$-th token of $y$, and $y_{<t}$ indicating the tokens preceding $y_t$.

\vspace{4pt}\noindent\textbf{Parameter Efficient Fine-Tuning.}
Fine-tuning all parameters of LLMs is time-consuming and resource-intensive.
To alleviate this challenge, Parameter-Efficient Fine-Tuning (PEFT) \cite{peft, p_tuning,prompt_tuning} optimizes a smaller set of parameters, significantly reducing computational requirements while still achieving commendable performance.
LoRA \cite{lora} is a typical PEFT algorithm, which keeps the LLM weights frozen and decomposes the updating weights into trainable low-rank matrices. 
The optimizing objective of LoRA can be formulated as follows:
\begin{equation}
\max_\Theta\sum_{(x,y)\in\mathcal{Z}}\sum_{t=1}^{|y|}\log\left(P_{\Phi_0+\Delta\Phi(\Theta)}(y_t|x,y_{<t})\right),
\label{eq:lora}
\end{equation}
where LoRA introduces parameters $\Theta$, which are smaller in size in comparison to the original LLM parameters $\Phi_0$.
\section{Large Language-Recommendation Assistant (LLaRA)}
\begin{figure*}[t]
\centering
\includegraphics[width=0.8\textwidth]{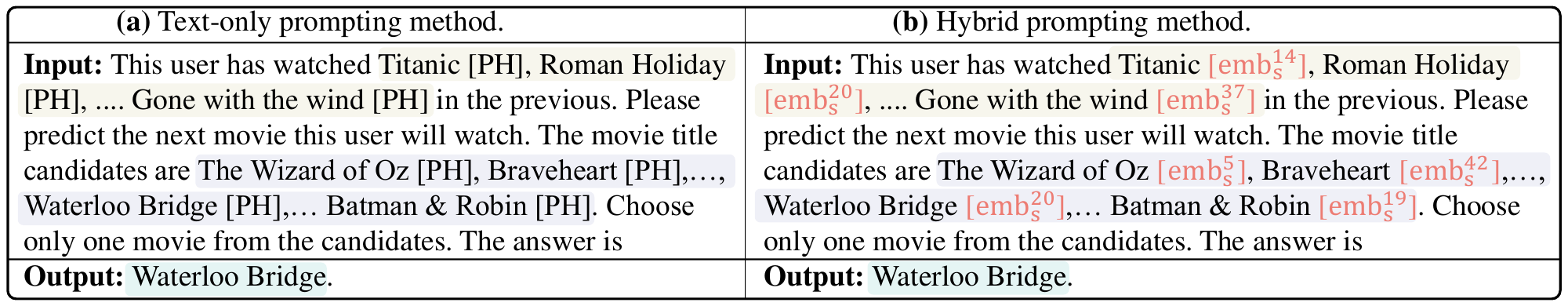}
\vspace{-4pt}
\caption{Illustration of text-only and hybrid prompting method. (a) Text-only prompting represents items with the combination of the textual token and a placeholder token. (b) Hybrid prompting represents items with the integration of the textual token and the behavioral token. Note that <PH> indicates a special placeholder token, reserved for substitution by the behavioral token $\text{<}emb_s^i\text{>}$ throughout the progressive learning procedure. }
\label{prompt_example}
\vspace{-4pt}
\end{figure*}

To incorporate the behavioral patterns learned by traditional recommenders into LLMs, we propose an end-to-end framework, Large Language-Recommendation Assistant (LLaRA), as depicted in Figure \ref{framework}.
Specifically, beyond the text-only prompting, it exploits a hybrid prompting to align the behavioral representations, as derived from recommendation systems, with the language space of LLMs.
It then employs curriculum learning --- first focusing on text-only prompting, then progressively transitioning to hybrid prompting.
This progressive strategy enables the LLM to familiarize the recommendation mechanism and internalize the behavioral knowledge of conventional recommenders.
We now delve into the detailed architecture and training paradigm of LLaRA.

\subsection{Item Representation}
\textbf{{Textual Token Representation.}}
Textual features of items, such as titles and descriptions, are the key to harnessing the commonsense knowledge inherent in LLMs.
Formally, for an item $i$ with the text metadata $txt_i$, we obtain its textual tokens $\text{<}\mathbf{emb_t^i}\text{>}$ as follows:
\begin{equation}
    \text{<}\mathbf{emb_t^i}\textbf{>}=\textbf{LLM-TKZ}(txt_i),
\label{eq:txt}
\end{equation}
where $\textbf{LLM-TKZ}(\cdot)$ presents the LLM tokenizer and word embedding layer, encapsulating the process of transforming textual metadata into token representations.
Such textual token representations of items, residing within the language space, are inherently compatible with LLMs.

\vspace{5pt}
\noindent \textbf{{Behavioral Token Representation.}}
In parallel, conventional sequential recommender models, such as GRU4Rec \cite{GRU4Rec}, Caser \cite{Caser}, and SASRec \cite{SASRec}, effectively capture sequential patterns within ID-based item embeddings after training on the historical interaction data.
Formally, for item $i$, its ID-based representation learned by the conventional recommendation model is expressed as:
\begin{equation}
    \mathbf{e_s^i}=\textbf{SR-EMB}(i;\Theta_{e}),
\end{equation}
where $\textbf{SR-EMB}(\cdot)$ is the function that generates the item embedding with the sequential recommender $\textbf{SR}$  parameterized by $\Theta_{e}$, and $\mathbf{e_s^i}\in\mathbb{R}^{d}$ is the $d$-dimensional representation of item $i$.

In contrast to the item-aware texts that can be naturally inserted into the prompt and easily interpreted by LLMs, the ID-based item representations might be incompatible with the text nature of LLM prompts.
Consequently, we view the ID-based representations as a distinct modality, separate from textual data.
To bridge the modality gap, it is essential to map the ID-based representation space of recommenders into the language space of LLMs.
This alignment allows LLMs to interpret and leverage the behavioral knowledge distilled by conventional recommenders.

To facilitate the alignment, we introduce a specialized module, SR2LLM, as illustrated in Figure \ref{projector}.
Specifically, we project the ID-based item representation $\mathbf{e_s^i}$ into the LLM space with a trainable projector $\mathbf{Proj}$ (\ie two-layer perceptions).
This process results in the generation of a behavioral token representation, $\text{<}\mathbf{emb_s^i}\text{>}$, formalized as:
\begin{equation}
    \text{<}\mathbf{emb_s^i}\text{>}=\mathbf{Proj}(\mathbf{e_s^i};\Theta_{p}),
\end{equation}
with $\Theta_{p}$ as the parameters of the trainable projector.

\vspace{5pt}
\noindent{\textbf{Hybrid Token Representation.}}
Upon acquiring the textual tokens $\text{<}\mathbf{emb_t^i}\text{>}$ and the behavioral token $\text{<}\mathbf{emb_s^i}\text{>}$ for item $i$, we proceed to integrate these two components.
This integration facilitates a comprehensive description of item $i$, effectively combining the distinct yet complementary aspects captured by each token:
\begin{equation}
\text{<}\mathbf{emb_c^i}\text{>}=\textbf{Concat}(\text{<}\mathbf{emb_t^i}\text{>},\text{<}\mathbf{emb_s^i}\text{>}).
\label{eq:hybrid}
\end{equation}

\subsection{Hybrid Prompt Design}
\noindent{\textbf{Text-Only Prompting.}}
For converting sequential interaction data into training data suitable for LLM instruction tuning, our initial approach adopts a straightforward method known as text-only prompting.
This approach represents items via textual metadata within the prompts, as illustrated in Figure \ref{prompt_example}a.
The input prompts $x$ encompass several key elements:

\textbf{(1) Task Definition:} a clear description of the sequential recommendation task (\eg ``predict the next movie this user will watch'').

\textbf{(2) Interaction Sequence:} the sequence of historical user-item interactions (\eg ``Titanic \text{<}PH\text{>}, Roman Holiday \text{<}PH\text{>}, \ldots, Gone with the wind \text{<}PH\text{>}'').

\textbf{(3) Candidate Set:} the set of candidate items, from which the LLM is to generate responses to the given task (\eg ``The Wizard of Oz <PH>, Braveheart \text{<}PH\text{>}, \ldots, Waterloo Bridge \text{<}PH\text{>}, \ldots, Batman \& Robin \text{<}PH\text{>}'').

Within the input prompt, each item is represented using the textual tokens followed by a placeholder token.
Additionally, the output $y$ comprises the textual tokens corresponding to the next item with which the user will engage (\eg ``Waterloo Bridge'').

\vspace{5pt}
\noindent{\textbf{Hybrid Prompting.}}
To incorporate behavioral insights captured by recommender models into the prompts, we devise a hybrid prompting method, as exhibited in Figure \ref{prompt_example}b.
Formally, we consider a user $u$ with a historical sequence of interactions involving items denoted as $h_1, h_2, \ldots, h_n$.
The user is presented with a set of candidate items, represented as $\mathbb{C}=\{c_1, c_2, \ldots, c_m\}$, from which the user may select the next item of interest.
Thus the three primary components of hybrid input prompts $x$ are transformed correspondingly as follows: 
\begin{enumerate}[leftmargin=*]
    \item \textbf{Task Definition:} identical to the text-only prompting method, which describes the sequential recommendation task in text.
    \item \textbf{Interaction Sequence with Hybrid Item Representations:} the sequence of historical user-item interactions, represented as $\text{<}\mathbf{emb_c^{h_1}}\text{>},\text{<}\mathbf{emb_c^{h_2}}\text{>}, \ldots , \text{<}\mathbf{emb_c^{h_n}}\text{>}$ (\eg Titanic $\text{<}\mathbf{emb_s^{14}}\text{>}$, Roman Holiday $\text{<}\mathbf{emb_s^{20}}\text{>}$, \ldots, Gone with the wind $\text{<}\mathbf{emb_s^{37}}\text{>}$).
    \item \textbf{Candidate Set with Hybrid Item Representations:} the set of item candidates represented with the integration of textual and behavioral tokens as $\text{<}\mathbf{emb_c^{c_1}}\text{>},\text{<}\mathbf{emb_c^{c_2}}\text{>},\ldots, \text{<}\mathbf{emb_c^{c_m}}\text{>}$, from which the LLM is expected to generate responses (\eg The Wizard of Oz $\text{<}\mathbf{emb_s^5}\text{>}$, Braveheart $\text{<}\mathbf{emb_s^{42}}\text{>}$,\ldots, Waterloo Bridge $\text{<}\mathbf{emb_s^{20}}\text{>}$,\ldots, Batman \& Robin $\text{<}\mathbf{emb_s^{19}}\text{>}$).
\end{enumerate}
This approach utilizes a fusion of textual and behavioral tokens, as formulated in Equation \eqref{eq:hybrid}, to represent items.
This contrasts with the text-only prompts, which rely solely on textual tokens as outlined in Equation \eqref{eq:txt}, thereby enriching the prompt with a more comprehensive understanding of user-item interactions.

Our hybrid prompt design facilitates integration of textual metadata and ID-based item embeddings sourced from a well-trained recommender model.
This design addresses the limitations of prompts that rely exclusively on either ID-based or textual data, thereby generating more accurate recommendations.

\subsection{Curriculum Prompt Tuning}
Considering the design of LLMs, which predominantly train on data in text, the task of comprehending modalities --- behavioral tokens distilled from recommender models --- presents a notable challenge. 
While the text-only prompting aligns closely with the LLMs' training and is thus more readily assimilated, the hybrid prompting, representing a deviation from typical language data, introduces a more complex task.

Drawn inspiration from curriculum learning \cite{curriculm_learning}, which emphasizes the importance of training the model from simple to more challenging learning tasks, we design a curriculum prompt tuning scheme in LLaRA.
In general, the tuning process begins by focusing on the more straightforward prompting method --- text-only prompting method.
This initial phase allows the model to establish a fundamental grasp of the sequential recommendation task. 
Subsequently, we gradually introduce the hybrid prompting method that incorporates behavioral tokens, thereby elevating the complexity of the tuning process.
This step-wise strategy ensures that the model is not overwhelmed by the complex task.
Ultimately, our LLM-based recommender will be fully integrated with the hybrid item representation.
This entire learning trajectory is shown as the gradient-colored rectangle in Figure \ref{framework}.

Formally, this learning process can be articulated through the subsequent stages, corresponding point-to-point with the three pivotal phases of curriculum learning.

\vspace{5pt}
\noindent \textbf{(1) Complexity Assessment:}
The initial step of curriculum learning is to assess the complexity of each task. 
In LLaRA, the task complexity is highly related to the integration of behavioral tokens in the hybrid prompt design.
Therefore, we define the easy and hard learning tasks, where the easy task adopts the sequential data reformatted into the text-only prompts as depicted in Figure \ref{prompt_example}a, whereas the hard task employs the data reformatted into the hybrid prompts as elucidated in Figure \ref{prompt_example}b.
Specifically, the loss function of the easy task can be formulated as:
\begin{equation}
L_{easy}(x^e, y^e) = - \sum_{t=1}^{|y^e|}\log\left(P_{\Phi_0+\Delta\Phi(\Theta)}(y^e_t|x^e, y^e_{<t})\right),
\label{eq:obj_easy}
\end{equation}
where $(x^e, y^e)$ is the text-only prompts shown in Figure \ref{prompt_example}a.
Besides, the loss function of the hard counterpart can be formulated as:
\begin{equation}
L_{hard}(x^h,y^h) = - \sum_{t=1}^{|y^h|}\log\left(P_{\Phi_0+\Delta\Phi(\Theta)+\Theta_p+\Theta_{e}}(y^h_t|x^h,y^h_{<t})\right),
\label{eq:obj_hard}
\end{equation}
where $\Theta_p$ and $\Theta_{e}$ are the parameters of the projector and the embedding layer of the conventional sequential recommender, respectively, and $(x^h, y^h)$ represents the hybrid prompt in Figure \ref{prompt_example}b.

\vspace{5pt}
\noindent \textbf{(2) Scheduler Formulation:}
After acquiring the learning objectives of the easy and hard tasks in Equation \eqref{eq:obj_easy} and \eqref{eq:obj_hard}, respectively, we can formulate the curriculum scheduler by transferring from the easy task to the hard task gradually in the training process.
Specifically, we denote $p(\tau)$ as the probability of learning the hard task at training time $\tau$, while $1-p(\tau)$ is the probability of the easy task, correspondingly. 
Naturally, $p$ should be small at the beginning and gradually increase in the learning process, which can be formulated in a continuous manner: 
\begin{equation}
    p(\tau)=\frac{\tau}{T}\quad (0\le \tau \le T),
\end{equation}
with the total training time accumulated to $T$.

\vspace{5pt}
\noindent \textbf{(3) Training Execution: }
To strike a balance between efficiency and efficacy, we conduct LoRA tuning as introduced in Equation \eqref{eq:lora} for the LLM, while training the projector at the same time.
Formally, we define the indicator function:
\begin{equation}
    \mathbb{I}(\tau) = 
    \begin{cases}
    1, & \text{learning hard task (\emph{w.p.} $p(\tau)$)}  \\
    0, & \text{learning easy task (\emph{w.p.} $1-p(\tau)$)} 
    \end{cases}.
\end{equation}
Therefore, the learning objective of LLaRA evolves from the easier task to the harder task:
\begin{equation}
\begin{split}
\min_{\Theta,\Theta_p}
\sum_{(x,y)\in\mathcal{Z}}(\left(1 - \mathbb{I}\left(\tau\right)\right) \ L_{easy}\left(x,y\right) +\mathbb{I}\left(\tau\right) \ L_{hard}\left(x,y\right)).
\end{split}
\end{equation}

This gradual learning process effectively facilitates the injection of an additional modality, thereby actualizing the hybrid prompting method.
By adopting the curriculum prompt tuning strategy, we ensure a seamless transition from the model's initial understanding of textual metadata to its eventual comprehension of more complex ID-based item embeddings from traditional recommenders.
This strategy not only acquaints LLMs with the recommendation mechanism, but also enhances LLMs with the behavioral knowledge encapsulated in the sequential recommenders.
\section{Experiments and Results}
In this section, we evaluate our proposed framework LLaRA on three real-world datasets, and compare it with several baselines, including traditional sequential recommender models and LLM4Rec models.
Additionally, we carry out two ablation studies to demonstrate the substantial enhancements brought about by the hybrid prompting method and curriculum prompt tuning strategy of LLaRA. 
Furthermore, we present case studies to explicitly show our advantages over baselines.
To validate the superiority of our framework, we will showcase it by answering research questions as follows.
\begin{itemize}[leftmargin=*]
    \item \textbf{RQ1: } How does LLaRA perform compared with traditional sequential recommender models and LLM-based methods?
    \item \textbf{RQ2: } How does our hybrid prompting perform in comparison to other forms of item representation in prompt design?
    \item \textbf{RQ3: } How does our curriculum learning scheme measure against other modality injection methods?
\end{itemize}

\subsection{Experimental Settings}
\begin{table}[t]
\centering
\small
\caption{Statistics of Datasets. }
\vspace{-12pt}
\label{tab:data_statis}
\begin{small}
\begin{tabular}{lrrr}
\toprule
Dataset& MovieLens & Steam & LastFM \\
\midrule
\# Sequence & 943 & 11,938 & 1,220 \\ 
\# Item & 1,682 & 3,581 & 4,606 \\ 
\# Interaction & 100,000 & 274,726 & 73,510 \\ 
\bottomrule
\end{tabular}
\end{small}
\vspace{-12pt}
\end{table}

\begin{table*}[t]
\centering
\caption{The Results of LLaRA compared with traditional sequential recommender models and LLMs-based methods. Bold and underlined indicate the best and the second-best performance, respectively. *($p$-value $<< 0.05$).}
\vspace{-5pt}
\label{tab:RQ1_res}
\small
\begin{tabular}{lccccccc}
\toprule
\multicolumn{2}{c}{\multirow{2}{*}{}}                                          & \multicolumn{2}{c}{MovieLens$^*$}                                                  & \multicolumn{2}{c}{Steam$^*$} & \multicolumn{2}{c}{LastFM}\\ 
\multicolumn{2}{c}{}                                                           & \multicolumn{1}{c}{ValidRatio} & HitRatio@1    & \multicolumn{1}{c}{ValidRatio} & HitRatio@1    & \multicolumn{1}{c}{ValidRatio} & HitRatio@1\\ 
\midrule
\multicolumn{1}{l}{\multirow{3}{*}{Traditional}} & GRU4Rec                     & \multicolumn{1}{c}{1.0000}     & 0.3750  & \multicolumn{1}{c}{1.0000}     & 0.4168  & \multicolumn{1}{c}{1.0000}     & 0.2616 \\ 
\multicolumn{1}{l}{}                             & Caser                       & \multicolumn{1}{c}{1.0000}     & 0.3861  & \multicolumn{1}{c}{1.0000}     & 0.4368 & \multicolumn{1}{c}{1.0000} & 0.2233  \\ 
\multicolumn{1}{l}{}                             & SASRec          & \multicolumn{1}{c}{1.0000}     & 0.3444  & \multicolumn{1}{c}{1.0000}     & 0.4010 & \multicolumn{1}{c}{1.0000} & 0.2233 \\ 
\midrule
\multicolumn{1}{l}{\multirow{4}{*}{LLM-based}}         & Llama2                     & \multicolumn{1}{c}{0.4421}  &    0.0421        & \multicolumn{1}{c}{0.1653}     & 0.0135 & \multicolumn{1}{c}{0.3443} & 0.0246\\ 
\multicolumn{1}{l}{}                             & GPT-4                       & \multicolumn{1}{c}{0.9895}     & 0.2000        & \multicolumn{1}{c}{0.9798}     & 0.3626 & \multicolumn{1}{c}{1.0000} & 0.3770 \\  
\multicolumn{1}{l}{}                             & MoRec                         & \multicolumn{1}{c}{1.0000}           & 0.2822& \multicolumn{1}{c}{1.0000}           & 0.3911 & \multicolumn{1}{c}{1.0000} & 0.1652\\ 
\multicolumn{1}{l}{}                             & TALLRec                     & \multicolumn{1}{c}{0.9263}     & 0.3895 & \multicolumn{1}{c}{0.9840}     &0.4637 & \multicolumn{1}{c}{0.9836} & 0.4180\\ 
\midrule
\multicolumn{1}{l}{\multirow{3}{*}{Ours}}        & LLaRA (GRU4Rec)             & \multicolumn{1}{c}{0.9684}     & \underline{0.4421}& \multicolumn{1}{c}{0.9975}     &\underline{0.4924} & \multicolumn{1}{c}{0.9836} & \underline{0.4344}\\ 
\multicolumn{1}{l}{}                             & LLaRA (Caser)               & \multicolumn{1}{c}{0.9684}     & \textbf{0.4737} & \multicolumn{1}{c}{0.9966} & 0.4874 & \multicolumn{1}{c}{0.9918} & \underline{0.4344}\\ 
\multicolumn{1}{l}{}                             & LLaRA (SASRec)              & \multicolumn{1}{c}{0.9684}     & \underline{0.4421} & \multicolumn{1}{c}{0.9975}     &\textbf{0.4949} & \multicolumn{1}{c}{1.0000} & \textbf{0.4508}\\ 
\bottomrule
\end{tabular}
\vspace{-5pt}
\end{table*}

\subsubsection{Datasets. }
\begin{itemize}[leftmargin=*]
    \item \textbf{MovieLens} \cite{movielens} is a commonly-used movie recommendation dataset that contains user ratings and movie titles.
    \item \textbf{Steam} \cite{SASRec} encompasses user reviews for video games on the Steam Store, in addition to game titles.
    \item \textbf{LastFM} \cite{lastfm}, collected from the Last.fm online music platform, includes user-artist listening relationships and the names of artists.
\end{itemize}

Given that tuning LLMs is more time-consuming than training traditional recommenders, we choose the MovieLens100K dataset for our experiment to ensure that the dataset size remains manageable.
Regarding the Steam dataset, we initially eliminate users with fewer than 20 reviews, aligning with the processing method employed for MovieLens.
Then, we randomly select a third of the users and a third of the games, maintaining their interactions to derive a dataset of a moderate size.
For all three datasets, we arrange sequences chronologically and divide the data into train, validation, and test subsets at a ratio of 8:1:1.
This partitioning approach guarantees that subsequent interactions do not appear in the training data, thereby circumventing any potential information leakage \cite{data_leakage}.
Detailed statistics of the datasets are provided in Table \ref{tab:data_statis}.
Moreover, we retain the last 10 interactions as the historical sequence, padding sequences with fewer than 10 interactions.

\subsubsection{Implementation Details. }
We select Llama2-7B \cite{llama2} as the LLM backbone.
To ensure the flexibility of our textual interface, the instruction format for training and testing is randomly sampled from several prompts.
Our implementations for conventional recommenders follow \cite{DROS}, employing the Adam optimizer, with a learning rate of 0.001, an embedding dimension $d$ of 64, and a batch size of 256. 
Furthermore, we conduct a grid search in [1e-3, 1e-4, 1e-5, 1e-6, 1e-7] for the coefficient of L2 regularization.
To mitigate the impact of randomness, we report the average outcomes of five runs using different random seeds.
For all methods related to LLMs, each experiment is trained for a maximum of 5 epochs, with a batch size of 128.
We employ a warm-up strategy for the learning rate, initiated with 1/100 of the maximum learning rate, and adjust it over steps using a cosine scheduler. 

\subsubsection{Evaluation Metrics. }
For each sequence, we randomly select 20 non-interacted items to construct the candidate set, ensuring the inclusion of the correct subsequent item.
LLaRA and other baseline models aim to identify the correct item from this candidate set, and their performance is evaluated using the HitRatio@1 metric.
With appropriate prompting, LLM-based recommenders can generate a single candidate item as required. 
As for traditional models, we select the candidate item with the highest probability as the prediction.
Meanwhile, since LLaRA employs a generative paradigm for prediction, which may yield invalid responses such as nonsensical words or items outside the candidate sets, we introduce an additional metric --- valid ratio. 
It quantifies the proportion of valid responses (\ie items in the candidate set) across all sequences, serving as a measure of the models' capability of instruction following.

\subsection{Performance Comparison (\textbf{RQ1})}
In this section, we compare LLaRA against both traditional and LLM-based baselines, taking into account metrics of both HitRatio@1 and valid ratio on MovieLens, Steam, and LastFM datasets, to showcase the effectiveness and robustness of LLaRA.

\subsubsection{Baselines}
\begin{itemize}[leftmargin=*]
    \item Traditional Sequential Recommenders: \textbf{GRU4Rec} \cite{GRU4Rec}, \textbf{Caser} \cite{Caser}, and \textbf{SASRec} \cite{SASRec}, are RNN-based, CNN-based, and attention-based sequential recommenders, respectively.
    \item LLM-based Models: (1) \textbf{Llama2} \cite{llama2} is a well-known open-source LLM released by Meta. (2) \textbf{GPT-4} \cite{GPT-4}, released by OpenAI, is a milestone of LLMs excelling in various tasks. (3) \textbf{MoRec} \footnote{For MoRec, we adopt BERT as the text encoder and SASRec as the recommender backbone, consistent with the officially provided implementation.} \cite{MoRec_or_IDRec} enhances the traditional recommenders by encoding item's modality features, such as text features.  (4) \textbf{TALLRec} \footnote{TALLRec predicts YES/NO for the target item, thus we adapt it to our setting -- selecting the next item from a provided candidate set.} \cite{TALLRec} conducts instruction tuning for LLMs on recommendation corpus. 
\end{itemize}

\begin{figure}[t]
\centering
\includegraphics[width=0.4\textwidth]{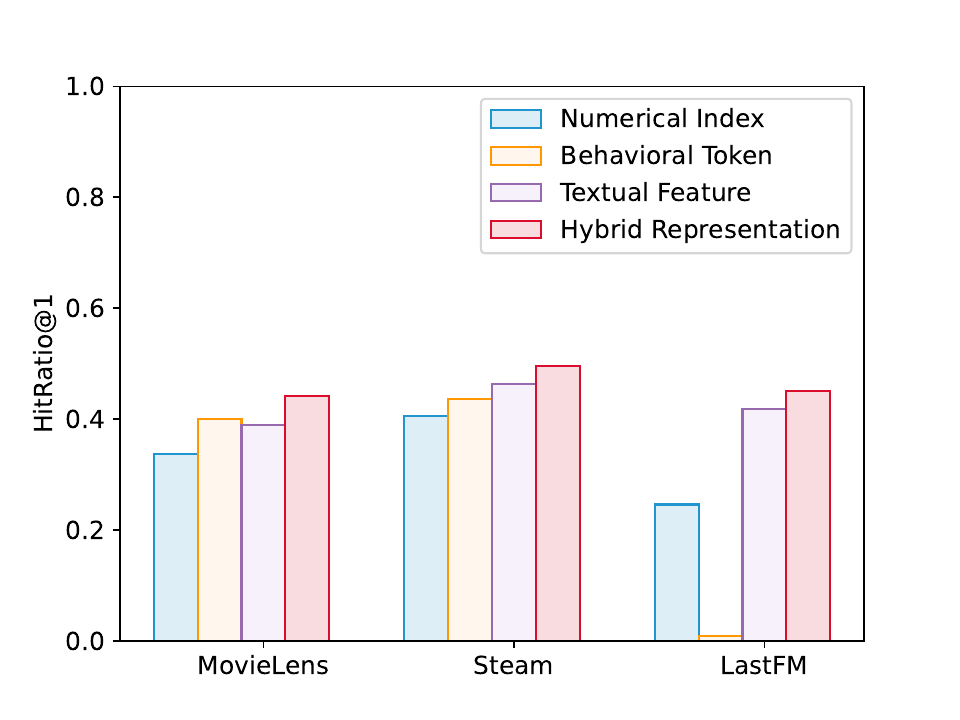}
\vspace{-12pt}
\caption{The performance comparison of different item representation methods (\ie numerical index, behavioral token, textual feature, and hybrid representation). 
The hybrid representation is adopted in LLaRA.}
\vspace{-10pt}
\label{ItemRepresentationResult}
\end{figure}

\subsubsection{Results}
We implement LLaRA framework on item embeddings derived from three traditional sequential recommendation baselines (\ie GRU4Rec, Caser, and SASRec).
Comparing LLaRA with the aforementioned baseline models, the results are shown in Table \ref{tab:RQ1_res} \footnote{Note that the relative improvement of GRU, Caser, and SASRec is calculated by the LLaRA implemented on the item embeddings derived from their corresponding models, while that of LLM-based methods is calculated by the LLaRA implemented on SASRec.}.
The observations can be summarized as follows.

(a) LLaRA outperforms all baselines on all three datasets. 
Specifically, it achieves the highest HitRatio@1 metric of 0.4737, 0.4949 and 0.4508 on MovieLens, Steam and LastFM, respectively.
This validates its effective integration of traditional sequential information with the extensive world knowledge and robust reasoning capabilities of LLMs.

(b) As for traditional sequential recommenders (\ie GRU4Rec, Caser, and SASRec), their HitRatio@1 scores are lower than those of LLaRA.
These models make predictions solely based on the behavioral patterns of users, without integrating any semantic information about items.
This highlights the importance of incorporating world knowledge about items into the recommendation process.

(c) When it comes to LLM-based methods, we can analyze them from two perspectives.
Firstly, the relatively poor performance of vanilla LLMs (\ie Llama2 and GPT-4) suggests that adapting LLMs to recommendation tasks is crucial for enhancing their performance in this domain.
Secondly, the LLM4Rec methods (\ie MoRec and TALLRec) show some improvements over the standalone LLM methods.
However, their recommendation ability, as denoted by the HitRatio@1 metric, is still lower than that of LLaRA.
MoRec overlooks the reasoning ability of LLMs, while TALLRec neglects to incorporate traditional sequential recommenders.
This highlights the need for a more comprehensive approach that combines the strengths of both LLMs and traditional recommendation models.

(d) LLaRA achieves a high validity ratio of over 95\% on all datasets, illustrating the model's instruction-following abilities when generating recommendations.
It's worth noting that all generative methods that incorporate LLMs might generate invalid answers.
For instance, Llama2, which serves as the backbone LLM of LLaRA, only achieves valid ratios of 0.4421, 0.1653, and 0.3443 on the MovieLens, Steam, and LastFM datasets, respectively.
Remarkably, LLaRA's significant improvement in valid ratios can be attributed to the fact that LLaRA has been instruction-tuned on the sequential recommendation task.

\subsection{Impact of Hybrid Item Representation (\textbf{RQ2})}
We conduct experiments to evaluate the item representation methods in sequential recommendation.

\begin{itemize}[leftmargin=*]
    \item \textbf{Numerical Index:} The items in the textual prompts are represented as numerical indices.
    \item \textbf{Behavioral Token:} The items are represented using behavioral tokens projected from the sequential recommender space, employing the identical projector architecture as LLaRA.
    \item \textbf{Textual Feature:} The items in the textual prompts are represented by their respective titles.
    \item \textbf{Hybrid Representation:} LLaRA proposes to represent items with the fusion of behavioral tokens and textual tokens.
\end{itemize}

The results are shown in Figure \ref{ItemRepresentationResult}, and we can observe that the item representation approach utilized by LLaRA surpasses other methods in terms of HitRatio@1 across all three datasets.
This not only corroborates the effectiveness of our innovative item representation method, but also illustrates the insufficiency of solely relying on semantic information (\ie textual metadata) or sequential information (\ie behavioral tokens).

Concerning numerical indices, no information is initially stored in LLMs for these indices.
The numerical indices are processed as plain text by LLMs, culminating in their separation into several tokens by the LLM tokenizer.
In the case of behavioral tokens, the LLM merely capitalizes on the distribution of the inputted behavioral embeddings, without eliciting the knowledge encapsulated within the LLM.
As for textual features, users' behavioral patterns are absent, allowing the LLM to solely infer the correlations among items in a user's historical interactions, guided by the background knowledge of these items preserved in the LLM.
In contrast, LLaRA integrates both world knowledge and sequential information, thereby improving performance in sequential recommendation.

\subsection{Impact of Curriculum Prompt Tuning (\textbf{RQ3})}
\begin{table}[t]
\centering
\caption{The HitRatio@1 of LLaRA compared with other learning strategies. CL denotes curriculum learning and bold indicates the best performance.}
\vspace{-8pt}
\label{tab:RQ3_res}
\small
\begin{tabular}{lccc}
\toprule
       & \multicolumn{1}{c}{MovieLens} & \multicolumn{1}{c}{Steam} & \multicolumn{1}{c}{LastFM}   \\
\midrule
Direct  & 0.4211 & 0.4899 & \textbf{0.4508}\\
Two-stage  & 0.4316 & 0.4840 & 0.4344\\
LLaRA (CL) & \textbf{0.4421} & \textbf{0.4949} & \textbf{0.4508}\\
\bottomrule
\end{tabular}
\vspace{-12pt}
\end{table}

This section delves into the development of an optimal learning strategy for modality integration, by comparing three schemes:

\textbf{(1) Direct Training:} The hybrid item representation is employed consistently during training.

\textbf{(2) Two-Stage Training}\textbf{:} The training process is split into two stages. Initially, Llama2 is fine-tuned on the easy task wherein the item representation is solely comprised of item titles.

\textbf{(3) LLaRA (CL):} LLaRA framework adopts a single-stage curriculum learning approach. Our curriculum learning strategy instructs the model to transition gradually from the basic text-only prompting to the hybrid prompting.

All training procedures encompass a total of five epochs for fair, while in the case of the two-stage method, the epoch number for the first and second stages is 2 and 3, respectively.

A careful analysis of the results, presented in Table \ref{tab:RQ3_res}, reveals that curriculum learning employed by LLaRA, consistently outperforms the other baseline methods across all datasets.
Specifically, the direct training method confounds the model with the hard task throughout the entire process, while the two-stage training approach fine-tunes Llama2 on the text-only and hybrid prompts in the first and second stages, respectively.
LLaRA starts from the easy task and progressively changes to the hard task utilizing a sampler to schedule the training process.
The improvement brought by this gradual learning method underscores the effectiveness of our curriculum prompt tuning scheme.

\subsection{Case Studies}
\begin{figure}[t]
\centering
\includegraphics[width=0.45\textwidth]{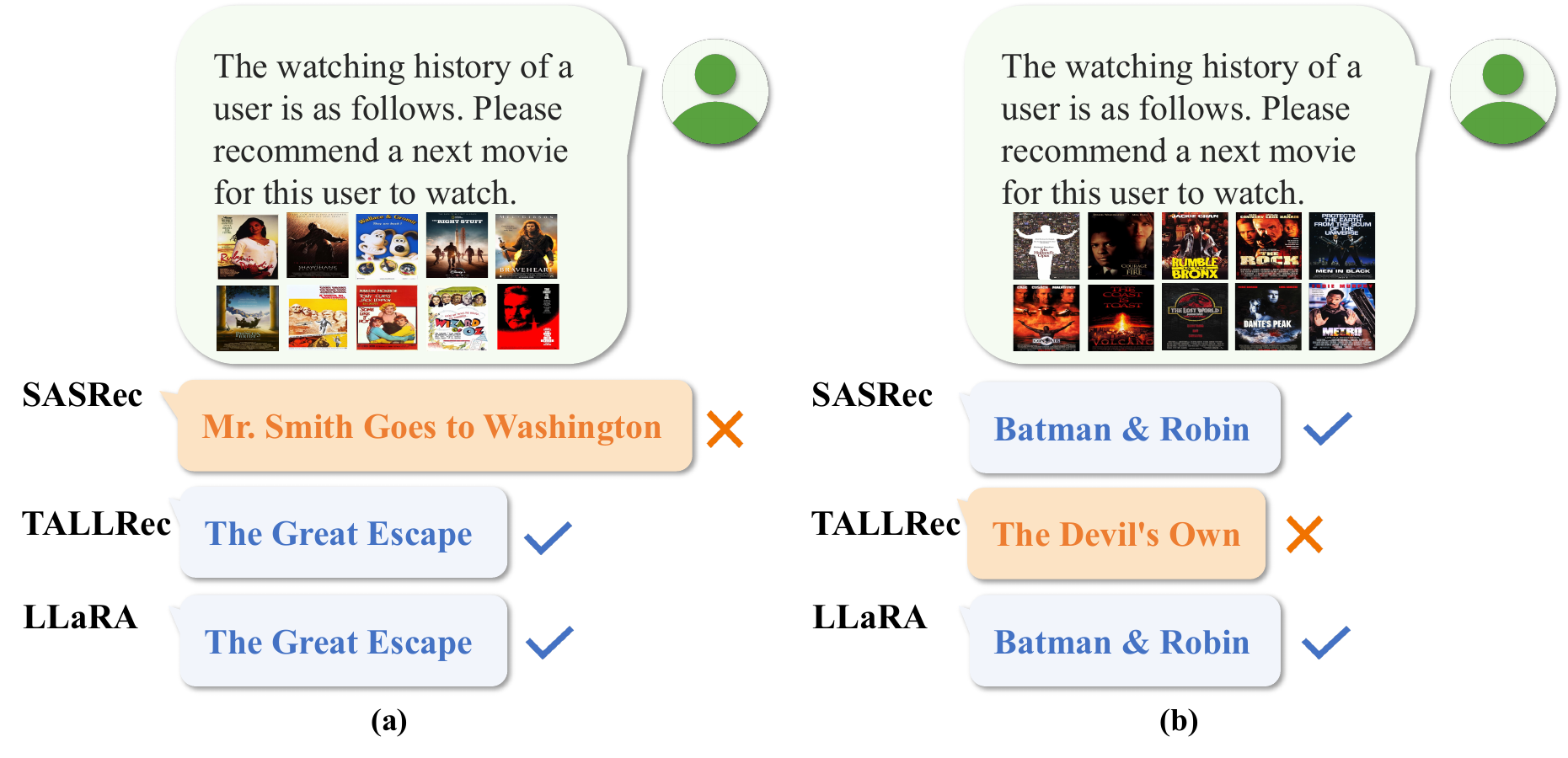}
\vspace{-5pt}
\caption{Case studies. (a) The user prefers adventure and war genres according to the viewing history. With the world knowledge about these movies, TALLRec and LLaRA correctly recommend ``The Great Escape''. (b) SASRec and LLaRA recommend ``Batman \& Robin'', according to the sequential behavioral patterns of users.}
\vspace{-8pt}
\label{case_study}
\end{figure}

We select two typical cases to analyze the impact of the world knowledge within LLMs, pertaining to items, as well as the behavioral patterns exhibited by users on the sequential recommendation task.
To illustrate these two factors, we choose the answers generated by three models, SASRec, TALLRec, and LLaRA.

\subsubsection{World Knowledge in LLMs}
For a user who sequentially watched ``Ruby in Paradise'', ``The Shawshank Redemption'', ``Wallace \& Gromit: The Best of Aardman Animation'', ``The Right Stuff'', ``Braveheart'', ``The Princess Bride'', ``North by Northwest'', ``Some Like It Hot'', ``The Wizard of Oz'', and ``The Hunt for Red October'', SASRec predicted the next film to be ``Mr. Smith Goes to Washington'', while TALLRec and LLaRA recommended ``The Great Escape''. The user's actual subsequent interaction was indeed ``The Great Escape'' as shown in Figure \ref{case_study}a.

We can observe that, the world knowledge about movies inherent in the LLM can be highly beneficial for the sequential recommendation, as demonstrated here.
The genres of films this user has watched include adventure (``The Princess Bride'', ``The Wizard of Oz'', ``North by Northwest'') and war (``Braveheart'', ``The Hunt for Red October'').
Since the LLM was capable of analyzing this user's watching history and understanding that the user has a preference for adventure and war genres, this insight allowed the LLM to correctly predict that the user would choose ``The Great Escape'' (a war adventure film) rather than ``Mr. Smith Goes to Washington'' (a political drama).
LLaRA, benefiting from the integration of the LLM's world knowledge, also forecasted the correct choice.

\subsubsection{Sequential Behavioral Patterns in Traditional Sequential Recommenders}
A user sequentially watched the following ten films: ``Mr. Holland's Opus'', ``Courage Under Fire'', ``Rumble in the Bronx'', ``The Rock'', ``Men in Black'', ``Con Air'', ``Volcano'', ``The Lost World: Jurassic Park'', ``Dante's Peak'', and ``Metro'' as shown in the Figure \ref{case_study}b.
TALLRec predicted the subsequent film to be ``The Devil's Own''; whereas both SASRec and LLaRA recommended ``Batman \& Robin'', which aligns with the user's actual interaction. 

TALLRec, based on background knowledge, may have inferred that the user prefers action, adventure, or thriller films over superhero movies.
``The Devil's Own'' is an action thriller, while ``Batman \& Robin'' is a superhero film.
However, SASRec, by analyzing the user's interaction history, unearthed sequential behavioral patterns and recommended the correct film.
LLaRA, due to the incorporation of information from SASRec, also predicted the correct answer.
This case illustrates that the sequential behavioral patterns of users hold substantial importance in sequential recommendation.

\section{Conclusion and Discussion}
In this paper, we introduce a novel framework, Large Language-Recommendation Assistant (LLaRA) that integrates traditional recommender models with LLMs, and transforms the sequential recommendation task into language modeling.
In particular, LLaRA adopts curriculum learning that gradually injects sequential patterns learned by traditional sequential recommenders into the tuning process of LLMs. 
Empirical results show that LLaRA outperforms all baseline models in sequential recommendation, demonstrating its effectiveness and promising performance.
Ablation studies underscore the essential role of both the hybrid prompting method and the curriculum prompt tuning strategy.

This work marks an initial step in transitioning from the traditional recommender models to a more sophisticated approach underpinned by LLMs and opens up new research possibilities.
It lays the groundwork by proposing an alignment mechanism to bridge conventional recommender models with LLMs.
In the future, researchers could continue to explore a unified recommendation framework, with natural language as the interface, for more complex and diverse recommendation scenarios.
We hope the development of LLaRA paves the way for a new era of personalized, integrated, and universal recommender systems.

\begin{acks}
This research is supported by the National Science and Technology Major Project (2023ZD0121102) and the National Natural Science Foundation of China (U21B2026, 62302321).
The work of Yancheng Yuan was supported by the Hong Kong Polytechnic University under grant P0045485.
\end{acks}

\bibliographystyle{ACM-Reference-Format}
\balance
\bibliography{LLaRA}

\end{document}